\documentclass[aps,twocolumn,showpacs]{revtex4}
\usepackage{graphicx}
\usepackage{amsfonts}
\begin{document}

\title{Dynamics of stick-slip  in  peeling of an adhesive tape}
\author{ Rumi De,$^1$
          Anil Maybhate,$^1$
     \footnote{Present address: Weill Medical College of Cornell University, New York, USA.}
          and G. Ananthakrishna$^{1,2}$,
     \footnote{Electronic mail: garani@mrc.iisc.ernet.in}
       }
\affiliation{$^1$ Materials Research Centre, Indian Institute of Science, Bangalore-560012, India\\
$^{2}$ Centre for Condensed Matter Theory, Indian Institute of
Science, Bangalore-560012, India}

\begin{abstract}
We investigate the dynamics of peeling of an adhesive tape
subjected to a constant pull speed. We derive the equations of
motion for the angular speed of the roller tape, the peel angle
and the pull force used in earlier investigations using a
Lagrangian. Due to the constraint between the pull force, peel
angle and the peel force, it falls into the category of
differential-algebraic equations requiring an appropriate
algorithm for its numerical solution. Using such a scheme, we show
that  stick-slip jumps emerge in a purely dynamical manner. Our
detailed numerical study shows that these set of equations exhibit
rich dynamics hitherto not reported. In particular, our analysis
shows that inertia has considerable influence on the nature of the
dynamics. Following studies in the Portevin-Le Chatelier effect,
we suggest a phenomenological peel force function which includes
the influence of the pull speed. This reproduces the decreasing
nature of the rupture force with the pull speed observed in
experiments. This rich dynamics is made transparent by using a set
of approximations valid in different regimes of the parameter
space. The approximate solutions capture major features of the
exact numerical solutions and also produce  reasonably accurate
values for the various quantities of interest.
\end{abstract}
\maketitle

\section{Introduction}

Peeling is a kind of fracture that has been studied experimentally
in the context of adhesion and is a technologically important
subject. Experimental studies on peeling of an adhesive tape
mounted on a cylindrical roll are usually in  constant pull speed
condition ~\cite{MB,HY1,Gandur,CGB,Bik,Kae}. More recently,
constant load experiments have also been reported
~\cite{Gandur,BC}. Early studies by Bikermann~\cite{Bik},
Kaeble~\cite{Kae} have attempted to explain the results  by
considering the system as a fully elastic object. This is clearly
inadequate as it ignores the viscoelastic nature of the glue at
the contact surface and therefore cannot capture many important
features of the dynamics. The first detailed experimental study of
Maugis and Barquins~\cite{MB} show stick-slip oscillations within
a window of pull velocity with decreasing amplitude of the pull
force as a function of the pull velocity. Further, these authors
report that the pull force shows sinusoidal, sawtooth and highly
irregular (chaotic as these authors refer to) wave patterns with
increasing velocities. More recently, Gandur {\it
et al.} have carried out a dynamical time series analysis of the
force waveforms, as well as those of acoustic emission signals and
report chaotic force waveforms at the upper end of the pull
velocities \cite{Gandur}. One characteristic feature of the
peeling process is that the experimental strain energy release rate
shows two stable branches separated by an unstable branch.
Stick-slip behavior is commonly observed in a number of systems
such as jerky flow or the Portevin-Le Chatelier (PLC)
effect~\cite{PLC}, frictional sliding~\cite{Poi}, and even
earthquake dynamics is thought to result from stick-slip of
tectonic plates~\cite{BK}. Stick-slip is characterized by the
system spending most part of the time in the stuck state and a
short time in the slip state, and is usually seen in systems
subjected to a constant response where the force developed in the
system is measured by dynamically coupling the system to a
measuring device. One common feature of such systems is that the
force exhibits ``negative flow rate characteristic" (NFRC).  Models
which attempt to explain the dynamics of such systems use the
macroscopic phenomenological NFRC feature as an input, although
the unstable region is not accessible. This is true for models
dealing with the dynamics of the adhesive tape as well. To the
best of our knowledge, there is no microscopic theory which
predicts the origin of the NFRC macroscopic law except in the case
of the PLC effect \cite{Anan82,Rajesh} (see below).

As there is a considerable similarity between the peeling of an
adhesive tape and the  PLC effect, it is useful to consider the
similarities in  some detail. The PLC effect refers to a type of
plastic instability observed when samples of dilute alloys are
deformed under constant cross head speeds \cite{KFG}. The effect
manifests itself in the form of a series of serrations in a range
of applied strain rates and temperatures. This feature is much
like the peeling of an adhesive tape. Other features common to
these two situations are: abrupt onset of the large amplitude
oscillations at low applied velocities with a gradually decreasing
trend and NFRC, which in the PLC effect refers to the existence of
negative strain rate sensitivity of the flow stress. In the case
of the PLC effect, the physical origin of the negative strain rate
sensitivity is attributed to the ageing of dislocations and their
tearing away from the cloud of solute atoms. Recently, the origin
of the negative SRS has been explicitly demonstrated as arising
from competing time scales of pinning and unpinning  in the
Ananthakrishna's model \cite{Anan82,Rajesh}. In the case of
adhesive tape, the origin of NFRC can be attributed to the
viscoelastic behavior of  the fluid. (Constant load and constant
load rate experiments are possible in the PLC also.) While simple
phenomenological models based on NFRC explain the generic features
of the PLC effect~\cite{Leb}, there appears to be some doubts if
the equations of motion conventionally used in the present case of
peeling are adequate to describe the velocity jumps
\cite{HY1,CGB}. Indeed, these equations of motion are singular and
pose problems in the numerical solutions.

Apart from detailed experimental investigation of the peeling
process, Maugis and Barquins \cite{MB}, have  also contributed
substantially to the understanding of the dynamics of the peeling
process. However, the first dynamical analysis is due to Hong and
Yue~\cite{HY1} who use an ``N" shaped function to mimic the
dependence of the peel force on the rupture speed. They showed
that the system of equations exhibits periodic and chaotic
stick-slip oscillations. However, the jumps in the rupture speed
are introduced {\it externally} once the rupture velocity exceeds
the limit of stability~\cite{HY2,CGB}. Thus, the stick-slip
oscillations are {\it not} obtained as a natural consequence of
the equations of motion.  Therefore, in our opinion the results
presented in Ref.~\cite{HY1} are the artifacts of the numerical
procedure followed. Ciccotti {\it et al.}~\cite{CGB} interpret the
stick-slip jumps as catastrophes. Again, the belief that the jumps
in the rupture velocity cannot be obtained from the equations of
motion appears to be the motivation for introducing the action of
discrete operators on the state of the system to interpret the
stick-slip jumps~\cite{CGB}, though they do not demonstrate the
correctness of such a framework for the set of equations.  Lastly,
there are no reports that explain the decrease in the amplitude of
the peel force with increasing pull speed as observed in
experiments. {\it As there is a general consensus that these
equations of motion correctly describe the experimental system, a
proper resolution of this question (on the absence of dynamical
jumps in these equations) assumes importance}.

The purpose of this paper is to show that the dynamics of
stick-slip during peeling can be explained using a
differential-algebraic scheme meant for such singular
situations~\cite{HLR} and demonstrate the rich dynamics inherent
to these equations. In what follows we first derive the equations
of motion (used earlier \cite{HY1}) by introducing an appropriate
Lagrangian for the system.   Then, we use an algorithm meant to
solve differential-algebraic equations \cite{HLR} and present the
results of our simulations for various parameter values.  One of
our major findings is that inertia has a strong influence on the
dynamics. In addition, following the dynamization scheme similar
to the one used in the context of the PLC effect \cite{Leb}, we
suggest that the peel force depends on the applied velocity. Using
this form of peel force  leads to the decreasing nature of the
magnitude of the pull force as a function of applied velocity. For
certain values of the inertia, we find canard type solutions.
These numerical results are captured to a reasonable accuracy
using a set of approximations valid in different regimes of the
parameter space. Even though, our  emphasis is on demonstrating
the correctness of these equations of motion and richness of the
inherent dynamics that  capture the qualitative features of   the
peeling process, we also attempt to make a comparison of the
experimental results mentioned above to the extent possible.
\begin{figure}
\includegraphics[height=3cm]{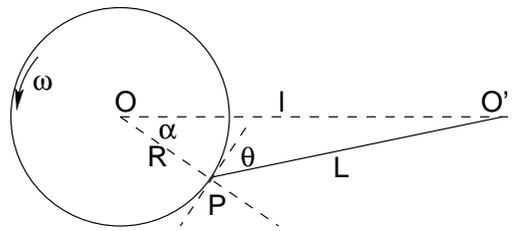}
  \caption{Schematic plot of experimental setup}
  \label{fig1}
\end{figure}

\section{EQUATIONS OF MOTION}

For the sake of completeness, we start by considering the geometry
of the experimental setup shown schematically in Fig. \ref{fig1}.
An adhesive roll of radius $R$ is mounted on an axis passing
through $O$ normal to the paper and is pulled at a constant
velocity $V$ by a motor positioned at $O'$ with a  force $F$
acting along $PO'$. Let the distance between $O$ and $O'$ be $l$,
and that between the contact point $P$ to $O'$ be $L$. The point
$P$ moves with a local velocity $v$  which can undergo rapid
bursts in the velocity during rupture. The force required to peel
the tape is usually called the force of adhesion denoted by $f$.
The two measured branches referred to earlier, are those of the
function $f$ in a steady state situation of constant pulling
velocity (i.e., there are no accelerations). The line $L$ makes an
angle $\theta$ with the tangent at the contact point $P$. The
point $P$ subtends an angle $\alpha$ at $O$, with the horizontal
line $OO'$. We denote the elastic constant of the adhesive tape by
$k$, the elastic displacement of the tape by $u$, the angular
velocity by $\omega$ and the moment of inertia of the roll by $I$.
The angular velocity itself is identified by $\omega = \dot \alpha
+ v/R$. The geometry of the setup gives $L\;{\rm cos}\,\theta =
-l\;{\rm sin}\,\alpha$ and $L\;{\rm sin}\,\theta = l\;{\rm
cos}\,\alpha - R$ which further gives, $L^2 = l^2 + R^2 -
2lR\;{\rm cos}\,\alpha$. The total velocity $V$ at $O^{\prime}$ is
then made up of three contributions~\cite{MB}, given by $V = v +
\dot u - \dot L,$ which gives
\begin{eqnarray}
v = V + \dot L - \dot u = V - R\;{\rm cos}\,\theta\;\dot\alpha - \dot u.
\label{v}
\end{eqnarray}
Following standard methods in mechanics, it is straightforward to
derive the equations of motion for $\alpha$ and $\omega$ by
considering $(\alpha,\dot\alpha,u,\dot u)$ as the generalized
co-ordinates. The corresponding Lagrangian of
the system can be written as
\begin{eqnarray}
{\cal L}(\alpha,\dot\alpha,u,\dot u) ={I\over2}[\omega(\alpha,\dot\alpha,u,\dot u)]^2 - {k\over2}u^2.
\label{LAGRAN}
\end{eqnarray}
We write the dissipation function as
\begin{eqnarray}
{\cal R} = \Phi(v,V)= \int f(v,V)dv, \label{DISSPN}
\end{eqnarray}
where $f(v,V)$ physically represents the peel force which we
assume is dependent on rupture speed as well as the pull speed
assumed to be derivable from a potential function $\Phi(v,V)$. The
physical origin of this is due to the competition between the
internal relaxation time scale of the viscoelastic fluid and the
time scale determined by the applied velocity~\cite{GL}. When the
applied velocity is low, there is sufficient time for the
viscoelastic fluid to relax. As we increase the applied velocity,
the relaxation of the fluid gets increasingly difficult and thus
behaves much like an elastic substance. The effect of competing
time scales is well represented by Deborah number~\cite{Bird}
which is the ratio of time scale for structural relaxation to the
characteristic time scale for deformation. Indeed, in the studies
on Hele-Shaw cell with mud as the viscous fluid, one observes a
transition from viscous fingering to viscoelastic
fracturing~\cite{Van} with increasing rate of invasion of the
displacing fluid.

\begin{figure}
\vbox{
\includegraphics[height=5cm,width=8cm]{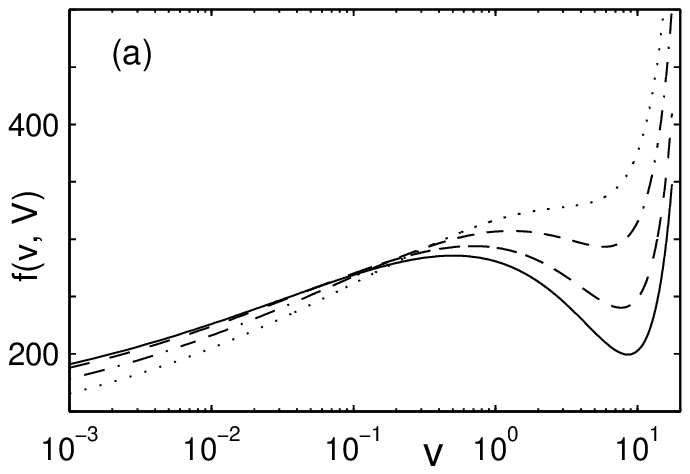}
\includegraphics[height=4cm,width=8cm]{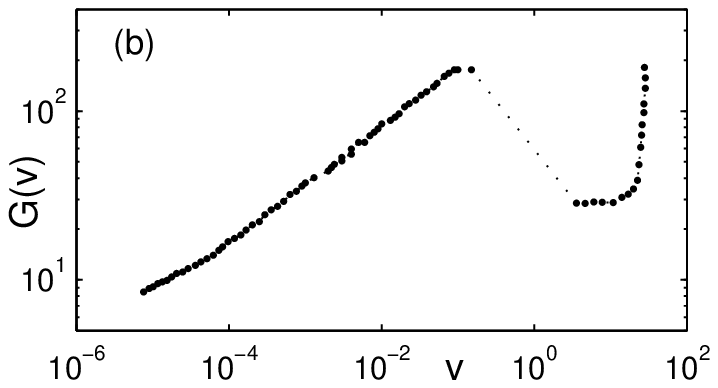}}
   \caption{(a) Plots of $f(v,V)$ as a function of $v$ ($x$ axis in log scale) for $V=1$
(solid curve), $V=2$ (dashed curve), $V=4$ (dashed and dotted curve),
               $V=6$ (dotted curve); see Eq. (\ref{fvV}). (b) Experimental 
strain energy release rate, $G(v)$ curve as in Ref. \cite{MB}. 
[Units of $f(v,V)$ is in N, $G(v)$ in J/m$^2$ and $v$, $V$ are in m/s ]}
  \label{fig2}
\end{figure}

As stated in the Introduction, the existing models do not explain
the decreasing amplitude of pull force. Similar feature observed
in the PLC serrations has been modeled  using a scheme referred
to as dynamization of the negative strain rate sensitivity (SRS)
of the flow stress $f(\dot \epsilon_P)$~\cite{Chihab,Leb}, where
$\dot \epsilon_p$ is the plastic strain rate. Based on arguments
similar to the preceding paragraph, they modify this function to
depend on the applied strain rate, $\dot \epsilon_a$, i.e.,
the negative SRS of the flow stress is taken to be $f(\dot
\epsilon_P, \dot \epsilon_a)$ such that the gap between the
maximum and the minimum of the function $f(\dot \epsilon_p, \dot
\epsilon_a)$ decreases with increasing $\dot \epsilon_a$.
Following this, we consider $f$ to depend on $V$ also, in a way
that the gap in $f$ decreases as a function of the pull speed $V$
(Fig. \ref{fig2}).

Using the Lagrange equations of motion,
\begin{eqnarray}
{d\over dt}\left({\partial{\cal L} \over {\partial\dot\alpha}}\right)-
{\partial{\cal L} \over {\partial\alpha}}+
{\partial{\cal R} \over {\partial\dot\alpha}}&=&0,
\label{MOTION1}
\\
{d\over dt}\left({\partial{\cal L} \over {\partial \dot u}}\right)-
{\partial{\cal L} \over {\partial u}}+
{\partial{\cal R} \over {\partial \dot u}}&=&0.
\label{MOTION2}
\end{eqnarray}
we obtain the same set of ordinary differential equations as in
Ref.~\cite{HY1} given by
\begin{eqnarray}
\dot\alpha &=& \omega - {v/R}, \label{FLOW1}\\
I \dot\omega &=& FR\;{\rm cos}\,\theta = -FR sin \alpha \simeq -FR
\alpha\label{FLOW2}\\
\dot F &=& k \dot u = k(V - v) - k\;{\rm cos}\,\theta (\omega R -
v),\label{flow3}\\
 &\simeq& k[ V-v + R \alpha \dot \alpha] \label{FLOW3},
\end{eqnarray}
with an algebraic constraint
\begin{eqnarray}
F(1-{\rm cos}\,\theta)-f(v,V) \simeq  F(1 + \alpha) - f(v,V)=0.
\label{CONSTR}
\end{eqnarray}
(The last equation results from the elimination of  two second
order equations for $\alpha$.) In Eqs.
(\ref{FLOW2}), (\ref{FLOW3}), and (\ref{CONSTR}) we have used $cos \theta
\simeq -sin \alpha \sim -\alpha$.  While Eqs. (\ref{FLOW1})-(\ref{FLOW3}) 
are differential equations, Eq. (\ref{CONSTR}) is an
algebraic constraint necessitating the use of
differential-algebraic  scheme to obtain the numerical solution
\cite{HLR}.

The fixed point of Eqs. (\ref{FLOW1}), (\ref{FLOW2}), (\ref{FLOW3}), and
(\ref{CONSTR}) is given by $\alpha=0, \omega=V/R, v=V, F=f(V,V)$. (For 
numerical solution, in the above equations we have actually
used $ sin \alpha $ in place of $ \alpha $.)  This point is stable
for $f'(V,V)>0$ and unstable for $f'(V,V)<0$. As $V$ is varied
such that the sign of $f'(V,V)$ changes from negative to positive
value, the system undergoes a Hopf bifurcation and a limit cycle
appears. The limit cycles reflect the abrupt jumps between the two
positive slope branches of the function $f(v,V)$.

\section{Algorithm}

The singular nature of these equations becomes clear if one were
to consider  the differential form of Eq. (\ref{CONSTR}) given by
\begin{eqnarray}
\dot v &=& {1\over f'(v,V)} \left[\dot F\;(1-{\rm cos}\,\theta) +
F\;{\rm sin}\,\theta\;\dot\theta\right], \label{vdot}
\\
& \simeq & [\dot F (1 + \alpha) + F \dot \alpha]/f^{\prime},
\label{vdotapprox}
\end{eqnarray}
where the prime denotes the derivative with respect to $v$.
Equations (\ref{vdot}) with Eq. (\ref{FLOW1}), (\ref{FLOW2}), and (\ref{flow3})
[or (\ref{FLOW3})] constitute the full set of evolution equations
for the vector $(\alpha,\omega,F,v)$. However, it is clearly
singular at points of extremum of $f(v,V)$, requiring an
appropriate numerical algorithm.

We note that Eqs.(\ref{FLOW1}), (\ref{FLOW2}), (\ref{flow3}), and (\ref{CONSTR}) can be written as
\begin{eqnarray}
M\dot{\bf X} = \phi({\bf X}),
\label{XDAE}
\end{eqnarray}
where ${\bf X} = (\alpha,\omega,F,v)$, $\phi$ is a vector function
that governs the evolution of ${\bf X}$ and $M$ is a singular
``{\it mass matrix}'' ~\cite{HLR} given by,
\begin{eqnarray}
M = \left(
\begin{tabular}{cccc}
1&0&0&0\\
0&1&0&0\\
0&0&1&0\\
0&0&0&0
\end{tabular}
\right).
\nonumber
\end{eqnarray}
Equation (\ref{XDAE}) is a differential-algebraic equation (DAE)
and can be solved using the so called singular perturbation
technique~\cite{HLR} in which the singular matrix $M$ is perturbed
by adding a small constant $\epsilon$ such that the singularity is
removed. The resulting equations can then be solved numerically
and the limit solution obtained as $\epsilon \rightarrow 0$. We
have checked the numerical solutions for $\epsilon$ values ranging
from $10^{-7}$ to $10^{-15}$ in some cases and the results do not
depend on the value of $\epsilon$ used as long as it is small. The
results presented below, however, are for $\epsilon = 10^{-7}$. We
have solved Eq. (\ref{XDAE}) using a standard variable-order solver, 
MATLAB ODE15S program.

We have parametrized  the form of $f(v,V)$ as
\begin{eqnarray}
f(v,V) &=& 400v^{0.35} + 110v^{0.15} + 130e^{(v/11)}- 2V^{1.5} \nonumber\\
&& -(415 - 45V^{0.4} - 0.35V^{2.15})v^{0.5} , \label{fvV}
\end{eqnarray}
to give values of the extremum of the peel velocity that mimic the
general form of  the experimental curves ~\cite{MB}. The measured
strain energy release rate $G(V)$ from stationary state
measurements is shown in Fig. 2(b).  The decreasing nature of the
gap between the maximum and minimum of $f(v,V)$ for increasing $V$
is clear from Fig. 2(a). [The values of $f(v,V)$ could not be
correctly determined as  $G(V)$ is in J/m$^2$ requiring more
details. However, the value  of $F_{max}$ is closer to Ref.
\cite{HY1} and the jumps in $v$ are similar to those in
experiments.] The reason for using the form given by Eq. (\ref{fvV})
is that the effects of dynamization are easily included through
its dependence on the pulling velocity while more complicated
terms are required to mimic completely the experimental curve (particularly 
the flat portion). However, we stress that the trend
of the results remains unaffected when the actual experimental
curve is used except for the magnitude of velocity jumps and the
force values.

\section{Results}

\begin{figure}[!t]
\includegraphics[height=4cm,width=8cm]{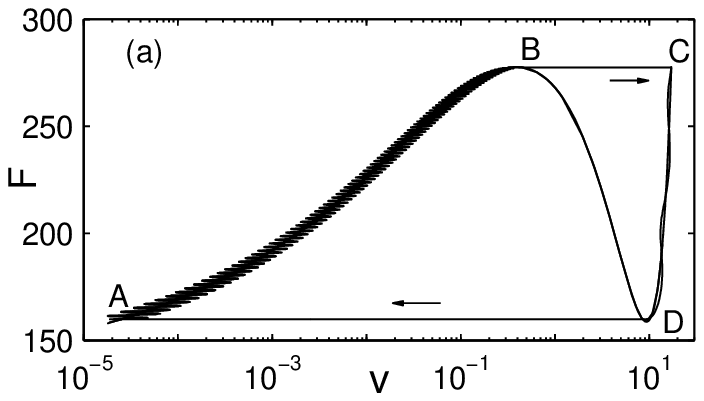}
\includegraphics[height=4cm,width=8cm]{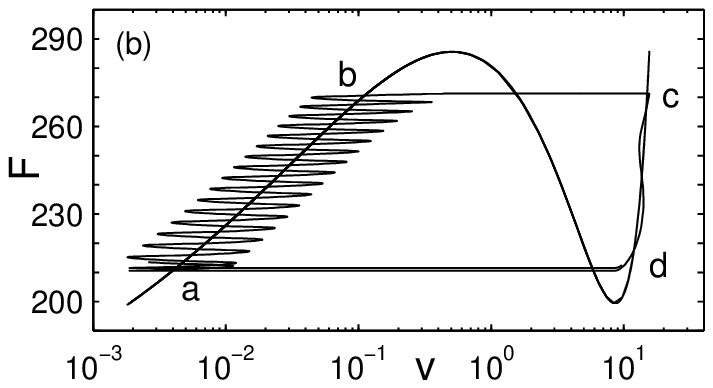}
\includegraphics[height=4cm,width=8cm]{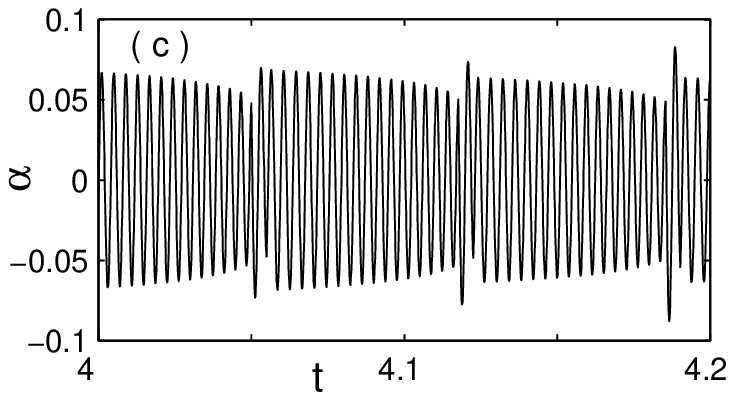}
\includegraphics[height=4cm,width=8cm]{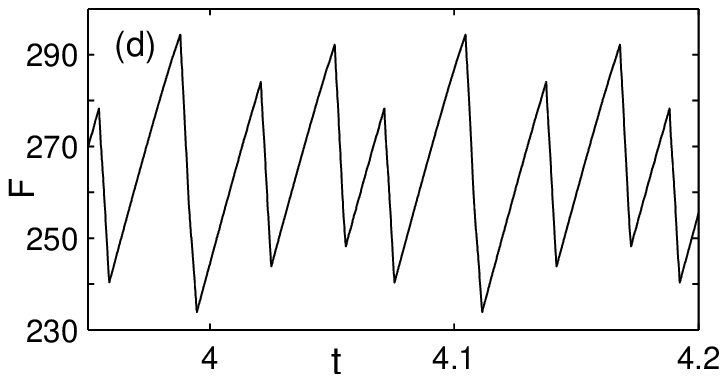}

\caption{(a) A typical phase space trajectory  in the $v-F$ plane for 
$V=0.4, I=10^{-5}$. The corresponding $f(v,V)$
is shown by a solid curve. (b) A phase space
trajectory in the $v-F$ plane for  $V=1.0$ and  $I=10^{-5}$. (c) A
plot of $\alpha(t)$ for $V=1$ and $I=10^{-5}$. (d) A plot of
$F(t)$ (period 4) for $V=2$ and $I=10^{-5}$. (Units of $v$, $V$ are in m/s, $F$ in $N$, $I$ in kg m$^2$ and $t$ in s.) }
\label{fig3}
\end{figure}

We have studied the dynamics of the system of equations for a wide
range of values of the parameters. We have found that transients
for some regions of parameters space take considerable time to die
out.  The results reported here are obtained after these long
transients are omitted.  These equations exhibit rich dynamics,
some even unanticipated. Here we report typical results for two
important parameters, namely the pull velocity $V$ (m/s) and the inertia
$I$ (kg m$^2$), keeping the elastic constant of the tape $k= 1000$ N/m, 
$R=0.1$ m and $l=1 $ m \cite{HY1}. The influence of $k$ will also
be mentioned briefly. ( Henceforth, we drop the units for the sake
of brevity.) We find that the observed jumps of the orbit in the
$v$-$F$ plane occur in a fully dynamical way. More importantly, we
find all the three possibilities namely, the orbit can jump when
it approaches the limit of stability, before or beyond that
permitted by $f(v,V)$. The dynamics can be broadly
classified into low, intermediate and high regimes of inertia.\\

(i) Low inertia. Here also, there are three regimes: low,
intermediate, and high pull velocity. \\

(a) Consider keeping inertia $I$ at a low value  (say $I= 10^{-5}$)
and  $V$ also at a low value (say, near the top, say $V =0.4$ ).
Here we observe regular saw tooth form for the pull force $F$. The
phase plot in the $F$-$v$ plane is as shown in Fig. 3(a). 
The corresponding function $f(v,V)$ is also shown by the
continuous curve. We see that the trajectory jumps almost
instantaneously from $B$ to $C$ on reaching  the maximum of
$f(v,V)$ (or from $D$ to $A$ when it reaches the minimum). The
system spends considerably more time on $AB$ compared to that on
$CD$. However, this feature of jumping of the trajectory at the
limit of stability is only true for small values of $I$ and when $V$
is  near the limit of stability. At slightly higher pull velocity,
say $V =1$, even for small $I$, say $I =10^{-5}$,  the jumps
occur even before reaching the top or bottom ( the points $B$ and $D$)
as can be seen from Fig. \ref{fig3}(b) for $V=1$. The small
amplitude high frequency oscillations seen in the phase plots [
Fig. \ref{fig3}(a), and \ref{fig3}(b)] on the branch $AB$ are due to the inertial
effect, i.e., finite value of $I$. These oscillations are better
seen  on the $\alpha(t)$ plot shown in Fig. 3(c).  For these values
of parameters, the system is aperiodic. \\

b) As we increase $V$, even as the saw tooth form of $F$ is
retained, various types of periodic orbits [period 4 shown in
Fig. \ref{fig3}(d) for $V =2$] as well as irregular orbits are seen.
In both cases (periodic as well as chaotic)  the trajectory jumps
from high velocity branch ($CD$) to the low velocity branch before
traversing the entire branch or sometimes going beyond the values
permitted by $f$. The value of $F$ at which the orbit jumps is
different for different cycles. For $I
=10^{-5}$, at high velocity, say $V=4$, the phase plot is periodic.\\

\begin{figure}[!t]
\includegraphics[height=5cm,width=8cm]{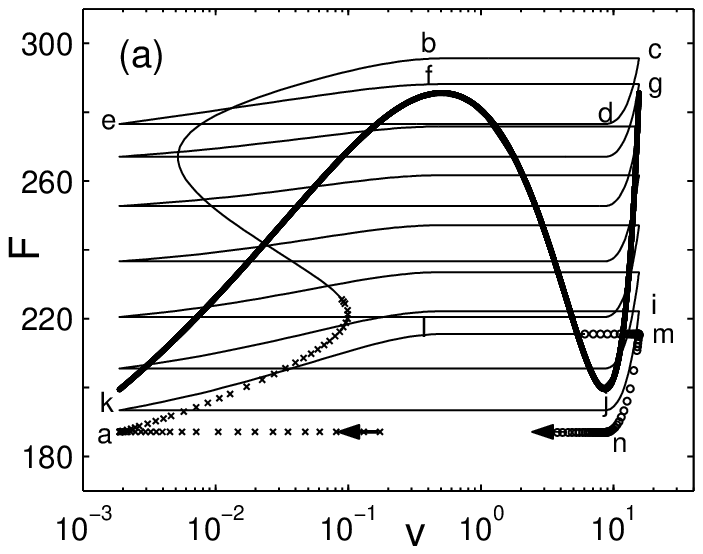}
\includegraphics[height=4cm,width=8cm]{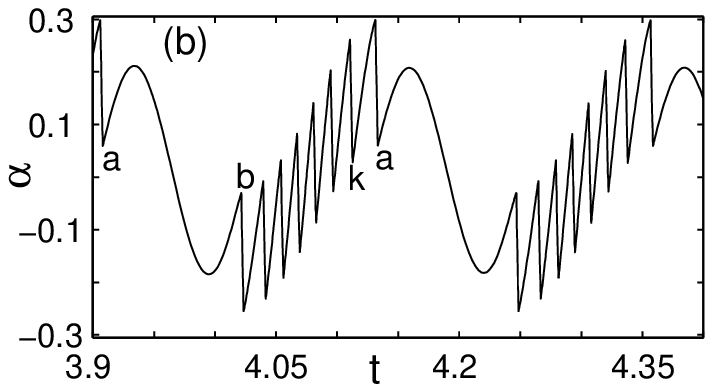}
\includegraphics[height=5cm,width=8cm]{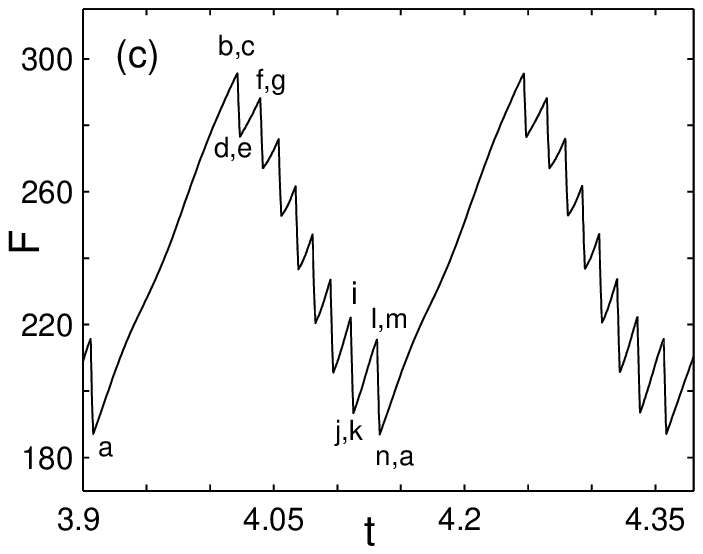}
\includegraphics[height=4cm,width=8cm]{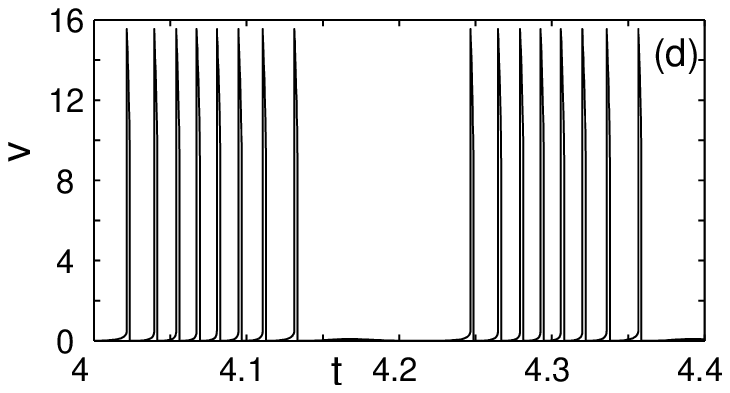}

\caption{(a) Phase space trajectory in  the $v-F$ plane for a
single cycle for $I =10^{-2}$ and $V=1$. The corresponding
$f(v,V)$ is shown by a thick solid curve. (b) Corresponding plots of $\alpha(t)$, (c) the pull force $F(t)$ (period 8) and (d) the peel velocity $v(t)$. (Units of $v$, $V$ are in m/s, $F$ in N, $I$ in kg m$^2$ and $t$ in s.) } \label{fig4}
\end{figure}

(ii) Intermediate and high inertia.\\
(a) As the results of small $V$ for intermediate and high inertia
are similar, we illustrate the results for $I =10^{-2}$ and $V
=1$. The $v-F$ phase plot, $\alpha, F$  and $v$ are shown in 
Fig. \ref{fig4}(a)-\ref{fig4}(d). Consider, Fig. \ref{fig4}(a) 
showing a typical phase
space trajectory for a single cycle. The corresponding function
$f(v,V)$ is also shown by the thick continuous curve. We see that
the maximum (and minimum) value of $F$ is larger (or smaller) than
that allowed by $f(v,V)$. [This feature holds when the inertia is
in the intermediate regime also, though the values of maxima (minima) 
of $F$ are not significantly larger (less) than $f_{max}$
($f_{min}$).] When the trajectory jumps from $AB$ to $CD$ at the
highest value of $F$ for the cycle, the trajectory stays on $CD$
for a significantly shorter time compared to the small inertia
case ($I=10^{-5}$) and jumps back to $AB$ well before $F$ has
reached the minimum of $f(v,V)$, i.e., $\Delta F$ is much smaller
than $f_{max} -f_{min}$. The pull force $F$ cascades down through
a series of back and forth jumps between the two branches till the
lowest value of $F$ for the cycle  is reached. Note that $F$ at
the point $n$ is less than $f_{min}$. For the sake of clarity, two
different portions of the trajectory are marked $abcdefg$ and
$ijklmna$ corresponding to the top and bottom regions of the plot.
The corresponding points are also identified on the $F(t)$ plot.
After reaching $n$, the orbit jumps to  $a$ on $AB$, the trajectory
decides to move up all the way till $F$ reaches a maximum value (larger 
than $f_{max}$, the point $b$) without jumping to the $CD$
branch. This part of $F$ as a function of time, which is  nearly
linear on $AB$, (i.e., the segment $ab$) displays a noticeable
sinusoidal modulation. The sinusoidal form is better seen in
$\alpha$ [Fig. \ref{fig4}(b)]. Note that the successive drops in
$F$ are of increasing magnitude. The jumps between the two
branches in the $v-F$ plane are seen as bursts of $v$ [Fig. \ref{fig4}(d)]. For these values of parameters,
the system is periodic. \\

\begin{figure}
\includegraphics[height=5cm,width=9cm]{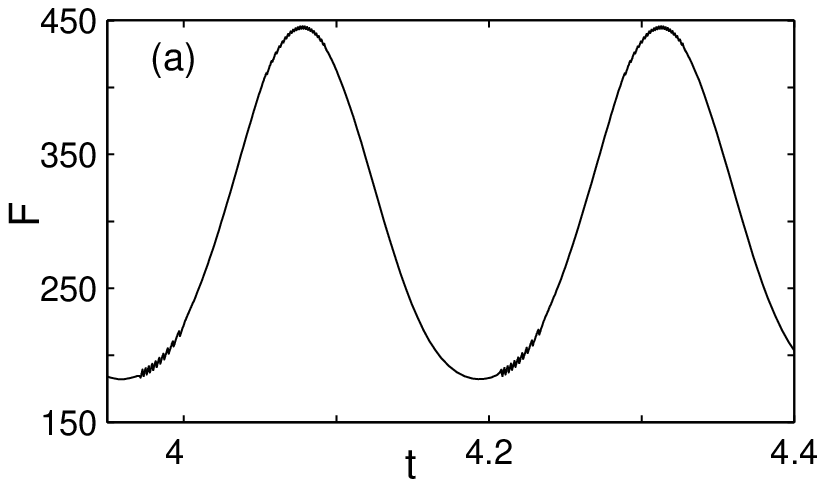}
\includegraphics[height=5.5cm,width=9cm]{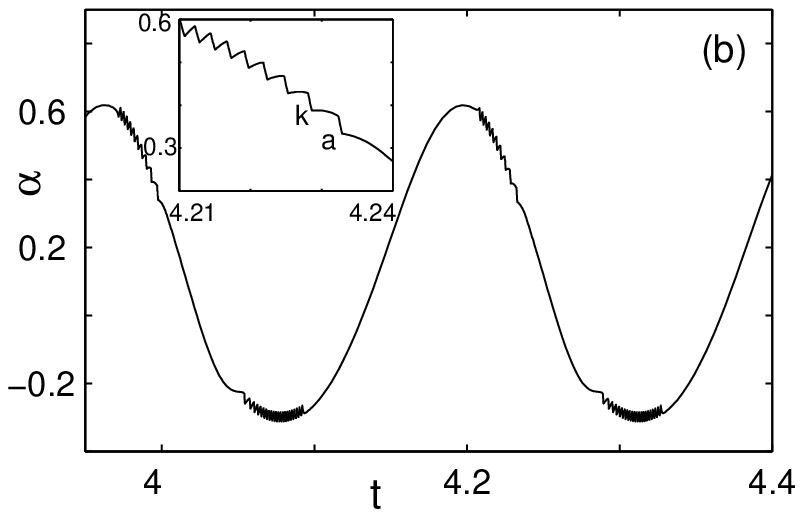}
\includegraphics[height=5.5cm,width=9cm]{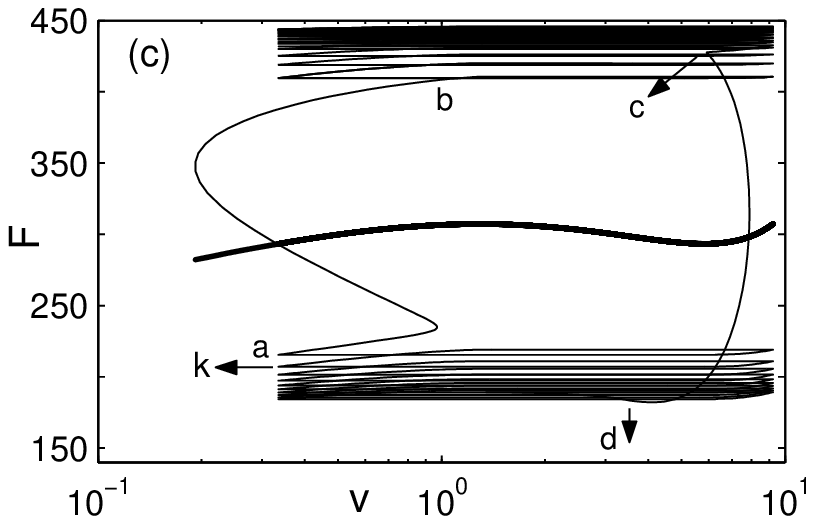}
\includegraphics[height=5cm,width=9cm]{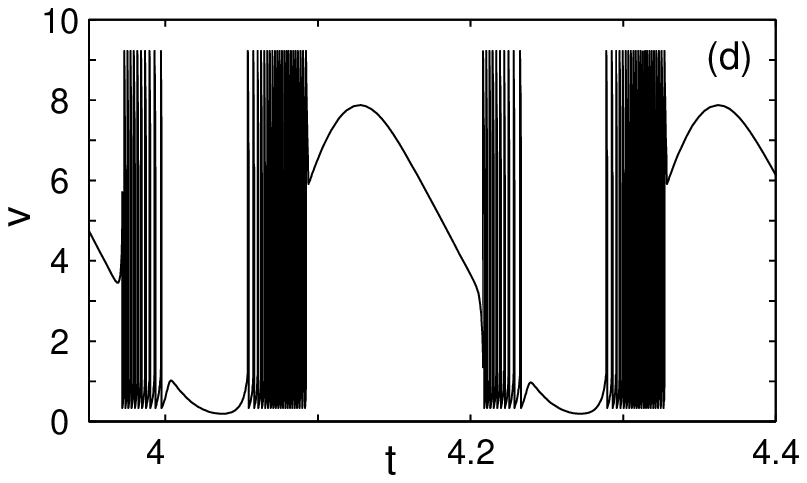}

\caption{(a) A plot of $F(t)$  for $V= 4$ and $I = 10^{-2}$. (b)
Corresponding plot of $\alpha(t)$. The inset shows an expanded
plot of decreasing trend of $\alpha(t)$. (c) Corresponding
plots of phase space trajectory that reflects the chaotic nature and (d) the
peel velocity $v(t)$. ($v$, $V$ are in m/s, $F$ in N, $I$ in kg m$^2$ and $t$ in s.)  } \label{fig5}
\end{figure}

(b) As we increase $V$, the sinusoidal nature of $F$ and $\alpha$
becomes more clear with its range becoming larger reaching a
nearly sinusoidal at $V=4$ for large $I$. [The range $ab$ in Fig. \ref{fig4}(c) expands. Compare Fig. \ref{fig5}(a).] The magnitude of
$\Delta F$ on the $CD$ branch for small $V$ and moderately or large
$I$, gradually decreases with increasing $V$.  The magnitude of
$\Delta F$  itself decreases as $I$ is increased. In the limit of
large $V$ and $I$, the drops in $ F$ and $\alpha$ become quite
small which are now located near the maxima and minima of these
curves. This is shown in Figs. \ref{fig5}(a) and \ref{fig5}(b). The sinusoidal
nature is now obvious even in $F(t)$ unlike for smaller $V$ and $I$
where it is clear only in $\alpha(t)$ for the low $v$ branch. Note
that for $V =4$, the nature of $f(v,4)$ is nearly flat. This
induces certain changes in the $v-F$ phase plot that are not
apparent in $F$ and $\alpha$. The jumps between the two branches
are now concentrated in a dense band at low and high values of
$F$. In this case, the maximum (minimum) value of $F$ is
significantly larger (less) than $f_{max}$ ($f_{min}$). These
rapid jumps between the branches manifest as jitter at the top and
bottom of $F$ and $\alpha$.

Unlike for small $V$ [Fig. \ref{fig4}(a)], the nature of the
trajectory in Fig. \ref{fig5}(c) is different. After reaching a
critical value of $F$ near the maximum value of $F$ (the point
$b$), the orbit spirals upwards and then descends down till another
critical value of $F$ (the point $c$) is reached. Having reached
$c$, the orbit monotonically comes down till $d$ where it jumps to
the AB branch. Beyond this point, it again spirals upwards till
the point $a$ is reached. Thereafter, $F$ monotonically increases
till $b$ is reached. The regions $ab$ and $cd$ are the regions
where $F$ shows a near sinusoidal form. The regions $bc$ and $da$
are the regions where the orbit jumps between the branches
rapidly. These manifest themselves as bursts of $v$ which tend to
bunch together almost into a band. [Compare Fig. \ref{fig4}(d)
with Fig. \ref{fig5}(d).] It is interesting to note that the jumps
between the two branches occur exactly at points where $df/dv =0$,
even when the maximum (minimum) of $F$ are higher (lower) than
that allowed by the stationary curve $f(v,V)$. The variables are
aperiodic for the set of parameters. The phase plots appear to be
generated by an effective $f(v,V)$ that is being cycled. [This
visual feeling is mainly due to the fact that jumps between the
branches still occur at the maximum and minimum of $f(v,V)$.]

The influence of $k$ is generally to increase the range of the
pull force  $F$ as can be easily anticipated and to decrease the
associated time scale.

It may be desirable  to comment on the similarity of the nature of
the force waveforms displayed by the model equations with those
seen in experiments. As mentioned in the introduction, apart from
qualitative statements on the waveforms in Ref. \cite{MB} (such as
periodic, sawtooth etc., which are seen in the model as well), it
should be stressed that there is a paucity of quantitative
characterization of the waveforms. In this
respect, the study by Gandur {\it et al.} \cite{Gandur} fills the gap to
some extent. These authors have carried out a dynamical analysis
of the time series for various values of the pull velocities (for
a fixed value of the inertia corresponding to their experimental
roller tape geometry). In order to compare this result, we  have
calculated the largest Lyapunov exponent for  a range of values of
$I$ and $V$. The region of chaos is in the domain of  small pull
velocities $V$ when $I$ is small. The maximum Lyapunov exponent
turns out to be rather high, typically around 7.5 bits/s in
contrast to the small values reported in Ref. \cite{HY1}. The
large magnitude of the positive exponent in our case can be traced
to the large changes in the Jacobian, as $df(v, V)/dv$ varies over
several order of magnitude($\sim 10^6$) as a function of the
peeling velocity and hence as a function of time. In contrast,
Hong {\it et al.} use an $N$ shaped curve where $df(v, V)/dv$ is
constant (and small) on both low and high $V$ branches.
However, these large values of Lyapunov exponents  are consistent
with rather high values obtained by Gandur {\it et al.}
\cite{Gandur} from time series analysis of the pull force. We also
find chaos for intermediate and high inertia in the region of high
velocities where the value of the Lyapunov exponent is small,
typically 0.5. The small value here again can be traced to the
small changes in $df(v, V)/dv$ at high velocities.

It must be mentioned that comparison with experiments is further
complicated  due to the presence of a two parameter family of
solutions  strongly dependent on both $I$ and $V$. Thus, the phase
diagram is complicated, i.e., the sequence of solutions
encountered in the $I$-$V$ plane as we change $V$ or $I$ or both
does not in general display any specific ordering of periodic and
chaotic trajectories (see Fig. 1 of Ref. \cite{Rajesh00}) usually
found in the well known routes to chaos. (For instance $2^n$
periods should be observed before the odd periods \cite{Licht}.)
Indeed, in our model, we find the odd periods 3,5,7 etc, on
increasing $V$ (or $I$), without seeing all the $2^n$ periods.
(These odd periods also imply chaos at parameter values prior to
that corresponding to these periods.) In view of this, a correct
comparison with experiments requires an appropriate cut in the
$I-V$ plane consistent with the experimental values of $I$ and $V$
even where they are given. However, as the values of $I$ are not
provided, full mapping of chaotic solutions is not possible. (We
also note that Gandur {\it et al.} \cite{Gandur} use a different
tape from that used in Ref. \cite{MB}, as is clear from the
instability range, leading additional difficulties in comparison.)

\begin{figure}
  \includegraphics[height=5cm,width=8cm]{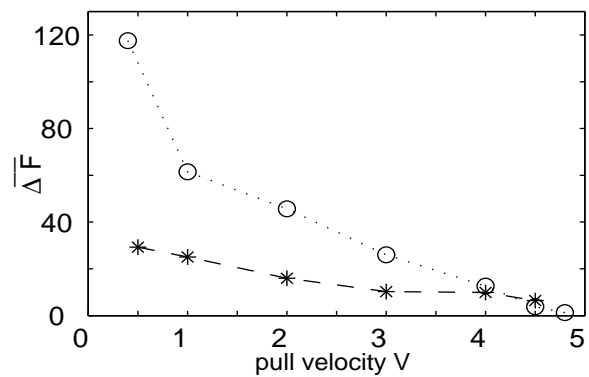}
   \caption{
The plot shows the mean force drop  $\overline{\Delta F}$ as a
function of the pull speed $V$, for two distinct values of $I$.
The dashed line corresponds to $I=10^{-2}$ while the dotted line
corresponds to $I=10^{-5}$. ($v$, $V$ are in m/s, $F$ in N, $I$ in kg m$^2$ and $t$ in s.) }
  \label{fig6}
\end{figure}

One quantitative result that can be compared with experiment is
the decreasing trend of the force drop magnitude. We have
calculated the magnitude of the force drops during stick-slip
phase as a function the pull velocity $V$ for both low ( $I
=10^{-5}$) and high ($I =10^{-2}$) inertia cases. Figure \ref{fig6} shows
 the monotonically decreasing trend of average
$\overline{\Delta F(t)}$  as $V$ is increased, for both small and
large $I$, a feature observed in experiments \cite{MB}. These two
distinct behaviors are  a result of the dynamization of $f(v,V)$
as in Eq.~(\ref{fvV}).

\begin{figure}
  \includegraphics[angle=-90]{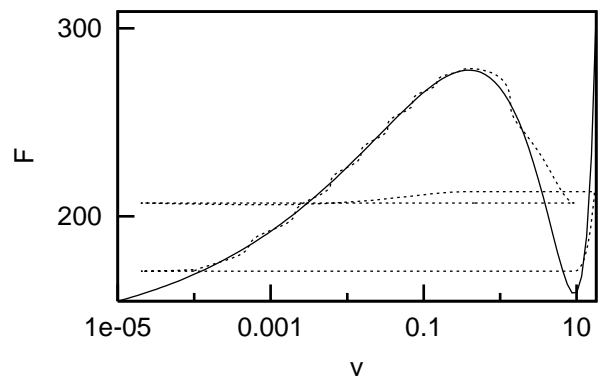}
\caption{A phase plot of canard type of solution in $v-F$ plane
for $V=0.4$ and $ I=10^{-3}$. ($v$, $V$ are in m/s, $F$ in N, $I$ in kg m$^2$ and $t$ in s.) }
  \label{fig7}
\end{figure}

Finally, as another illustration of the richness of the dynamics
seen in our numerical simulations, we show  in Fig. \ref{fig7}, a
plot of an orbit that sticks to unstable part of  the manifold
before jumping back to the $AB$ branch. Such solutions are known as
canards \cite{Canard}. Though canard type of solutions are rare,
we have observed them for high values of $I$ and low values of
$V$. In our case,  such solutions are due to the competition of
time scale due to inertia and that due to $v$. This again
illustrates the influence of inertia of the roll on the dynamics
of peeling.

It is clear that these equations exhibit rich and complex
dynamics. A few of these features are easily understandable, but
others are not. For instance, the saw-tooth form of $F$ for low
inertia and low pull velocity can be explained as resulting from
the trajectory sticking to stable part of $f(v,V)$  and jumping
only when it reaches the limit of stability. For these parameter
values, as the time spent by the system is negligible during the
jumps between the branches $AB$ and $CD$ (and vice versa), the
system spends most of the time on the branch $AB$ and much less on
$CD$ due to its steep nature. Then, from Eq. (\ref{FLOW3}), it is
clear that we should find a sawtooth form whenever the peel velocity
$v$ jumps across the branch to a value of $v$ larger  than the
pull velocity $V$.

However, several features exhibited by these system of equations
are much too complicated to understand. We first list the issues
that need to be explained.\\

(I) Small $I$.\\
(a) We find high frequency tiny oscillations superposed on the
linearly increasing $F$ [on the $AB$ branch or better seen in the
$\alpha$ plot Fig. \ref{fig3}(c)]. This needs to be understood.\\

(b) The numerical solutions show that the influence of inertia can
be important {\it even for small} $I$ and small $V$. For instance,
the jumps between $AB$ and $CD$ branches occur even before $F$ reaches
the extremum values of $f$.

(II) For intermediate and high values of inertia, for low $V$ case.\\
(a) We observe  several relatively small amplitude saw tooth form
of $F$ on the descending part of the pull force $F$. These appear
as a sequence of jumps between the two branches in the $v-F$ plane
which we shall refer to as the ``jumping mode". A proper estimate
of the magnitude of $\Delta F$ is desirable.

(b) In addition, there appears to be a critical value of $F$ for a
given cycle below which the return jumps from AB to CD stop and
one observes a
monotonically increasing trend in $F$ [$ab$ in Fig. \ref{fig4}(c)].\\

(III) High I and high $V$.

(a) The jumps between the branches occur at a very high frequency
[Fig. \ref{fig5}(c)] and now are located near the extremum values
of $F$ and $\alpha$. But these regions are separated by a stretch
where the orbit monotonically increases on the $AB$ branch and
monotonically decreases on the $CD$ branch. We need to elucidate the
underlying causes leading to the switching between the jumping
mode and monotonically increasing or decreasing mode.

(b) For large $V$, say $V=4$ and large $I$ (Fig. \ref{fig5}), the
extent of values of $F(t)$ range between $185$ and $450$ much
beyond $f(v,4)$ whose range is around 300. This feature is less
dominant for small $I$ and small $V$ case.

\section{APPROXIMATE ANALYSIS OF THE DYNAMICS}

As the dynamics is described by a coupled set of differential
equations with an algebraic constraint, the results are not
transparent. We first attempt to get insight into the complex
dynamics through some simple approximations valid in each of the
regimes of the parameters. Solution of these approximate equations
will require  appropriate initial values for the relevant
variables which will be provided from the exact numerical
solutions. Due to the nature of approximations, the results are
expected to capture only the trend and order of magnitudes of the
effects that are being calculated. But as we will show, even the
numbers obtained match quite closely with the exact numerical
results.

Our idea is to capture the dynamics through a single equation ( as
far as possible or at most two as in the high $I$ and $V$ case) by
including all the relevant time scales and solve the relevant
equation {\it on each branch.}  For this we note  that the
equation for $\alpha$ and $v$ play a crucial role as  the inertial
contribution appears only through Eqs. (\ref{FLOW1}) and (\ref{FLOW2})
and the time spent by the system is controlled by the equation for
$v$, Eq. (\ref{vdotapprox}). Using Eqs. (\ref{FLOW1}) and (\ref{FLOW2}),
we get
\begin{equation}
\ddot \alpha  =  - \frac{F(t) R\alpha}{I}  - \dot v/R. \label{genalpha}\\
\end{equation}
The general equation for $\alpha$ can be written
down by using Eq. (\ref{vdotapprox}), in Eq. (\ref{genalpha}), we get
\begin{eqnarray}
\ddot \alpha &=&-\frac{FR \alpha} {I} - \frac{[\dot F (1
+\alpha) + F\dot \alpha] }{Rf^{\prime}}, \label{genalpha1}\\
& \simeq &  -\frac{FR\alpha }{I} -\frac{ [\dot F + F
\dot\alpha]}{Rf^{\prime}}. \label{Airy1}
\end{eqnarray}
In obtaining Eq. (\ref{Airy1}), we have used $1 + \alpha \simeq 1$
which is valid except for high $I$ and high $V$.  Further, in most
cases, we can drop $R\alpha\dot\alpha $ as the magnitude of this
term is small and use $\dot F \simeq k(V-v)$. To be consistent we
use $F(t) \simeq F_{in} + k (V - v) t$. We note however that even
for high $I$ and high $V$ where $\alpha$ is not small, dropping $1
+\alpha$ and $R \alpha \dot \alpha$ causes only 10\% error.\\

\centerline {\bf Case I, small $I$}

On the low velocity branch $AB$,  as $v/R$ is small in Eq. (\ref{FLOW1}),
 we can drop $\dot v$ term in Eq. \ref{genalpha}.
Thus,
\begin{equation}
I\ddot {\alpha} \approx -F R \alpha. \label{approxalpha}
\end{equation}
Note that for the low inertia case, $sin \alpha \approx \alpha$
approximation is clearly justified [see Eq. (\ref{FLOW2})]. Using
this equation, we first get an idea of the relevant time scales as
$I$ is increased. \\

\centerline {\bf Case a}

Consider the low velocity branch $AB$ where the small
amplitude high frequency oscillations are seen on the nearly
linearly increasing part of $F$ [given by $F (t) = F_{min} + k (V
-v )t$, see for instance  Fig. \ref{fig3}(b)]. A rough estimate of
this time spent on this branch is obtained by $(f_{max} - f_{min})
/k V \sim t$. Using $f_{max} \sim 284$ and $f_{min} \sim 200$,
[from Fig. \ref{fig3}(b)], we get $t = 0.084$ (compared to the
correct value of $0.063$ which we shall obtain soon) which is much
larger than the period of the high frequency oscillation. Thus, we
could take the local value $F$ for the purpose of calculating the
period of the high frequency oscillation. Consider the orbit at
the lowest value of $F$ for which we can use $F_{min} \sim
f_{min}(v,1) \sim 200$. Then using Eq. (\ref{approxalpha}), the
frequency $\nu = {\sqrt (FR/I)}/2 \pi = 225$ for $I =10^{-5}$
which gives the period of oscillation $T = 4.44 \times 10^{-3}$.
This agrees very well with the exact numerical value   $T = 4.1
\times 10^{-3}$. This frequency decreases when the force reaches
the maximum value $F_{max} \sim f_{max}(v,1) \sim 284$ to $\nu
=261$ giving $T = 3.69 \times 10^{-3}$ which is again surprisingly
close to  the numerical value $3.72 \times 10^{-3}$. In the
numerical solutions, we find that the period gradually decreases
[see Fig. \ref{fig3}(c)]. This feature  is also easily recovered by
using $F = F_{min} + k (V-v) t$. This leads to an additional term
in the equation of motion for $\alpha$ in Eq. (\ref{approxalpha}),
\begin{equation}
I\ddot \alpha = - F(t) R\alpha/I= - [F_{min} + k(V-v) t] R\alpha/I, \label{Airy}
\end{equation}
where $t$ is the time required for $F$ to reach $F_{max}$ starting
from $F_{min}$. Here again  the $v$ term can be dropped.  If
$F_{min}$ was absent, the equation has the Airy's form. (Note
that for this case also we could assume $F_{min} \sim f_{min}$ and
$F_{max} \sim f_{max}$.) Though this equation does not have an
exact solution, we note that we could take $\alpha$ to have a
sinusoidal form with $2\pi\nu = \sqrt (FR/I)$ where $F$ is treated
as a slowly increasing parameter. (This assumption works quite
well.) The above equation captures the essential features of the
numerical solution. The numerical solution of Eq. (\ref{Airy}) ( as
also this representation ) gives the decreasing trend of the small
amplitude high frequency oscillations. (Note that the Airy
equation itself gives a decreasing amplitude \cite{Abro}.)

We note that Eq. (\ref{approxalpha}) is valid on the $AB$ branch where
$v$ is small even for high inertia and small $V$ case. Thus, we
may be able to recover the gross time scales using this equation.
Our numerical results show that as we increase the inertia,
$\alpha$ exhibits a sinusoidal form on the $AB$ branch [see Fig. \ref{fig4}(b)], although one full cycle is not seen. We note that
though the value of $\alpha$ is much larger than that for small
$I$, we can still use the above equations [Eq. (\ref{approxalpha})
and (\ref{Airy})]. On this branch $F$ increases from a value $
F_{min} \sim f_{min}$ to a maximum $F_{max} \sim f_{max}$. For
large $I=10^{-2}$ ( and $V=1$), we get a rough estimate of the
period by using the mean value of $F \sim 240$ in Eq.
\ref{approxalpha}. This gives a period $T = 0.128$ which already
agrees satisfactorily with the numerically exact value $T$ =0.11
considering the approximation used (i.e., using the mean $F$). A
better estimate can be obtained by using Eq. (\ref{Airy}).

For the high $I$ and $V$ case, Fig. 5 for $V=4$ shows that the
wave forms are nearly sinusoidal except for a jitter at the top
and bottom. For this case, $f(v,4)$ is nearly flat over the entire
range of values of $v$,  with a value $\sim$ 300. Here, even on
the AB branch, we can not ignore the $\dot v$ term in Eq. (\ref{genalpha}). However, one sees that as $v_{min} = 0.335$ and
$v_{max} = 1.25$ which suggest that to the leading order, we could
ignore the $\dot v$ term. This gives the period $T =0.115$. From
Fig. \ref{fig5}(b), considering only  the monotonically decreasing
part ($ab$), the value of $T/2=0.53$ read off from the figure
compares reasonably well with this value.

For the $CD$ branch, as $v$ is not small, the $\dot v/R$ term appears to
be important in Eq. (\ref{genalpha}). Some idea of when this term is
important  can be had by looking at the time scales arising from
inertia, namely,  $FR/I$ and the coefficient of the damping term,
$F/Rf^{\prime}$ in Eq. (\ref{Airy1}). Consider $V=1$ for $I =
10^{-5}$ and $10^{-2}$. The period obtained by assuming the mean
value of $F=240$ in $FR/I$ gives $4 \times 10^{-3}$ for $I
=10^{-5}$ compared to $0.128$ for $I = 10^{-2}$. These numbers can
be compared with the time scale $Rf^{\prime}/F$ which is 0.01 (where we 
have used $f^{\prime} \sim 25$ from numerical simulations
for $V =1$). This shows that for high inertia the damping
coefficient $F/Rf^{\prime}$ in Eq. (\ref{Airy1}) is important. We
will discuss this issue in more detail later.\\

\centerline{\bf Case b}

Now we focus on the origin of jumps between the branches. We note
that the jumps from $CD$ to $AB$ (or vice versa) occur only when the
peel velocity $v$ reaches a value where  $f^{\prime}(v,V) =0$.
This also means that the time scale on each branch, whether it
spends only a short time or not, is controlled by the equation for
$v$. However, clearly the influence of inertia needs to be
included.  Here we present an approximate equation for $v$ which
is valid in the various limits of the parameters:
\begin{eqnarray}
\dot v & = & [\dot F(t) (1+\alpha(t)) + F(t) \dot \alpha]/f^{\prime},
\label{fullvdot}\\
& \simeq & \frac {[ k (V-v )+ ( F_{in} + k (V-v)t ) \dot
\alpha)]}{f^{\prime}}, \label{approxvdot}
\end{eqnarray}
where the time $t$ is time spent on the branch considered ( low or
high $V$). In Eq. (\ref{approxvdot}), we have again used   $\dot F
\simeq k (V-v)$ and $F \simeq F_{in} + k(V-v)t$ with the same
approximation used in Eq. (\ref{Airy1}).

We now attempt to obtain correct estimates of the time spent by
the orbit on each branch starting with the least complicated
situation of the low inertia and small $V$. For this case, on the
low velocity branch, one can use the sinusoidal solution for
$\alpha$, namely $\alpha = \alpha_{in} sin ( 2 \pi \nu t + \phi)$,
where $\phi$ is a phase factor which also includes the
contribution arising from the jump as well and $2\pi\nu = \sqrt
(FR/I)$ with $F \simeq F_{in} + k V t$ . Both $\alpha_{in}$ and
$\phi$ needs to be supplied. Alternately, one can use Eq. (\ref{approxalpha})
 with Eq. (\ref{approxvdot}) for which we provide
$\alpha_{in}$ and $\dot \alpha_{in}$ at the point from the exact
numerical solutions. We stress that this procedure is {\it not}
equivalent to solving all the equations, as the only equation we
use is Eq. (\ref{approxvdot}) with the form of $\alpha$ already
determined from the equation for $\alpha$. [We note here that
though we have used the sinusoidal form of $\alpha$ along with the
initial conditions on $\alpha_{in},\phi$, it is simpler to supply
the initial conditions $\alpha_{in},\dot\alpha_{in}$ and use
Eq. (\ref{approxalpha}).] We note here that $f^{\prime}$ is a crucial
factor that determines the time at which the  orbit jumps from one
branch to the other. Equation (\ref{approxvdot}) needs to be
integrated from $v_{in}$ to $v_{f}$ that are determined by the
pulling velocity $V$, i.e., the form of $f(v,V)$.

\begin{figure}
  \includegraphics[height=6cm,width=8cm]{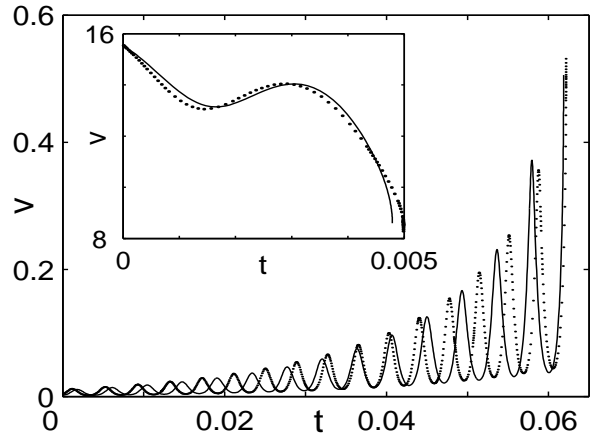}
  \caption{Comparison of approximate solution (continuous) with the
  numerically exact solution (dotted) of $v(t)$ for $I=10^{-5}$ and $V=1$ for the AB branch. The inset shows a similar comparison of $v(t)$ on
  the CD branch. ($v$, $V$ are in m/s, $I$ in kg m$^2$ and $t$ in s.)}
  \label{fig8}
\end{figure}

For the low $v$ branch $f^{\prime}$ term makes a significant
contribution for the time spent by the trajectory on $AB$. Indeed,
one can obtain the order of magnitude of the time spent by the
orbit on AB by using a crude approximation for $f^{\prime} (v, 1)
= -230 (0.5 -v)$. This can be easily integrated from $v = v_{in}
\sim 0. 0188$, to $v= v_{f} \sim 0.4$ which already gives $\Delta
t = 0.075$. This number is comparable to the numerically exact
value $0.063$. A correct estimate can be obtained by  using
$f^{\prime}$ from Eq. (\ref{fvV}) with the sinusoidal form $\alpha$
or Eq. (\ref{approxalpha}). (We have used $F_{in} = 211.5$ from the
numerical simulations for $V=1$  and $2 \pi \nu = \sqrt{[R
F(t)/I]}$ with $F = F_{in} + k (V -v) t$.) This gives nearly the
exact numerical value of $\Delta t = 0.063$. In fact, this
solution also captures the oscillatory growth nature of $v$ quite
accurately. The approximate form of $v(t)$ (continuous line) along
with the numerically exact solution ( dotted line) are shown in
Fig. \ref{fig8}. Using $\Delta t $ in $F = F_{in} + k (V-v) \Delta
t$ gives $\Delta F = 63$ and $F = 274.5$ which is in good
agreement with the exact numerical value of $F_{max} = 275$. It is
interesting to note that this value is much less than $f_{max} =
283$ [see Fig. \ref{fig3}(b)] or equivalently $\Delta F$ is less
than $f_{max} - f_{min}$, what is also observed in our exact
numerical simulation. The underlying mechanism  of jumping of the
orbit before $F$ reaches $f_{max}$ also becomes clear from the
analysis (Fig. \ref{fig8}). We note that the magnitude of the
oscillatory component  in $v$ grows till it reaches $v_{max}$
permitted by $f(v,1)$. Then, the orbit has to jump to $CD$. Thus,
the approximate solution gives an insight into the cause of the
orbit jumping even before $F$ reaches $f_{max}$ (for small I).

For the $CD$ branch also, the dominant term is  $f^{\prime}$.
Indeed, any reasonable function which has the same geometrical
form of $f$ shown in Fig. \ref{fig2} will give good results for
$\Delta t$. Using the correct form of $f^{\prime}$, we get $\Delta
t =0.005$ which is close to the exact result. This again gives
correct magnitude of $\Delta F = 72.5$.  In addition the nature of
the $v(t)$ obtained by this approximation is close to the  exact
numerical solution shown in the inset of Fig. 8.

\centerline {\bf Case II, intermediate and high $I$ and low $V$}

The most difficult feature of our numerical solutions to
understand is the dynamical mechanism leading to a series of drops
in the pull force seen on the descending branch of $F(t)$ for
intermediate and high values of inertia and for a range of $V$
values. Consider the high inertia and low $V$ case (say $I =
10^{-2}$ and $V=1$) shown in Fig. \ref{fig4}. As stated earlier,
there are two different issues that need to be understood here.
First, the series of small force drops $\Delta F$ and second the
monotonic increasing nature of $F$ on the $AB$ branch.

In this case, as already discussed, the coefficient of $\dot
\alpha$, namely,  $F/Rf^{\prime}$ term in Eq. (\ref{Airy1})
determines the time scale on $CD$, while on $AB$, the term
$FR\alpha/I$ dominates. Thus, the general equation valid for this
case is
\begin{equation}
\ddot \alpha \simeq  -\frac{FR\alpha }{I} -\frac{F
\dot\alpha}{Rf^{\prime}(v)}, \label{Airy2}
\end{equation}
where we use $F= F_{in} + k (V -v) t$. [Note that we have dropped
$k(V-v)$ term from Eq. (\ref{Airy1}) as this term does not have any
dependence on $\alpha$ or $\dot \alpha$.]

We start with the cascading effect. Consider the orbit when it is
at the highest value of $F_{in} = 295.6$ on the $CD$ branch for
which we can drop $FR\alpha/I$ term. As $f^{\prime}$ is a function
of $v$, and $F$ also depends on time, it appears that we need to use
coupled equations $\dot \alpha = - F\alpha /Rf^{\prime}$ with 
Eq. (\ref{approxvdot}). However, the numerical solution of these
equations show that one can make further approximation by taking
$f^{\prime}$ to be constant taken at $v= 15.54$  and $F=F_{in}$,
as the time spent on this branch is very small. The error in using
this approximation is within $10\%$. Indeed, using $\alpha_{in} =
-0.0304, \dot \alpha_{in} = -160 $ and numerically integrating 
Eq. (\ref{approxvdot}), along with Eq. (\ref{Airy2}) from $v_{in} =15.54$
to $v_f =8.7$ gives $\Delta t = 1.78 \times 10^{-3}$.  This
compares reasonably with the numerical value of $1.89 \times
10^{-3}$.  Using this we get $\Delta F = 19.4$ which compares well
with the numerically exact value 19.8. At this point the orbit
jumps  to the low velocity $AB$ branch (to the point $e$). Thus, as
$\Delta t$ is small,  for all practical purposes, we can ignore
the dependence of $f$ on $v$ and $F$ on $t$ and use $\alpha$ to be
an exponentially decreasing function for analytical estimates.
These analytical estimates already give reasonably accurate
numbers.

On the $AB$  branch, the dominant time scale is determined by
$FR\alpha/I$, and  we can use the approximate sinusoidal form in
Eq. (\ref{approxvdot}), or Eq. (\ref{approxalpha}) along with 
Eq. (\ref{approxvdot}) for the time evolution from the point $e$.
Integrating from $v_{in}= 0.0188$ to $v_f =0.4$ with the
appropriate initial values $\alpha_{in} = 0.239, \dot \alpha_{in}
= 11.9$ (or $\alpha, \phi$) and $F_{in} =276.53$, gives $\Delta t
= 0.016 $ which again compares very well with exact numerical
value $\Delta t =0.0164$. This gives $\Delta F = 11.61$. The
procedure for calculating the time spent by the orbit on $CD$ and $AB$
is the same and we find that successive values of $\Delta F$
increases which is again consistent with what is seen 
in Figs. \ref{fig4}(a) and \ref{fig4}(c).

\begin{figure}
\includegraphics[height=5cm,width=8cm]{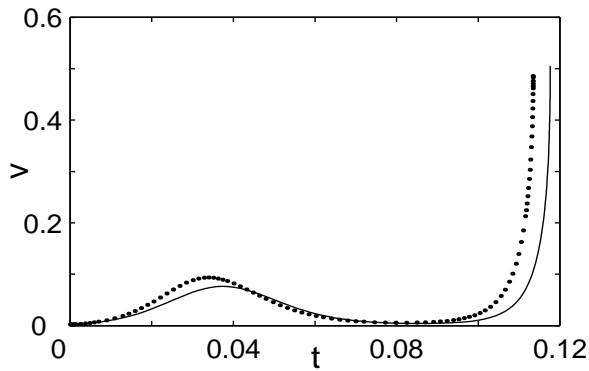}

   \caption{ Comparison of the approximate solution (continuous line)
   for $v(t)$ with numerically exact solution (dotted line)
for $V=1, I =10^{-2}$. ($v$, $V$ are in m/s, $I$ in kg m$^2$ and $t$ in s.) }
  \label{fig9}
\end{figure}

Continuing this  procedure, we find that a minimum value of $F =
186.95$ for the cycle is reached. Now consider  the time evolution
of $F$ on $AB$ that should lead to a monotonically increasing nature
as seen in the numerically exact solution. As this point
corresponds to the point at which the dynamics switches from the
jumping mode to the monotonically increasing nature of $F$ (i.e.,
the stretch $ab$), we discuss this in some detail.  For the point
$a$, we have used the initial condition $\alpha_{in} = 0.0599,
\dot \alpha_{in} = 9.7$ and integrating  Eq. (\ref{approxvdot}) and
Eq. (\ref{approxalpha}) (or the sinusoidal form of $\alpha$) from
$v= v_{min}=0.0188$ to $ v = v_{max} =0.4$ gives $\Delta t =
0.117$. This is nearly the value 0.114 obtained from the exact
numerical integration. This gives $\Delta F = 117$ and $F_{max}
=303.95$ which compares very well with the exact numerical value.
In addition, the growth form of $v$ obtained from this
approximation (continuous line)  agrees very well with that of
the exact numerical solution (dotted line) as shown in Fig. \ref{fig9}. The discrepancy seen in the figure can be reduced for
instance if we include the terms neglected in Eq. (\ref{genalpha1})
such as $\dot F$ and using $1+ \alpha$ in Eq. (\ref{fullvdot}).

Now we come to the crucial question. How  does the system know
that it has to go from $a$ to $b$,  while just during the previous
visit to the point $k$ on  $AB$ branch lead only to  a small
increase in $\Delta F$ [Fig. \ref{fig4}(a) and \ref{fig4}(c)] before 
jumping to $CD$?
To understand this, we recall that on $AB$, a sinusoidal solution is
allowed. First, one can notice a few differences in the initial
conditions between the point $a$ and $k$. For the point $k$, the
initial conditions taken from the exact numerical solution are
$\alpha_{in} = 0.298$ and $\dot \alpha_{in}= 18.3$. ( $F_{in}
=193.37$), while for the point $a$, $\alpha_{in} = 0.0589, \dot
\alpha_{in} = 9.7$. However, for $\alpha$  to begin  a sinusoidal
form, the initial value of $\dot \alpha = 18.3$ is much higher
than  the natural slope. The  local slope  for any sinusoidal form
is maximum when the variable is close to zero.  In Fig. \ref{fig4}(b),
 the sinusoidal form starts when $\alpha$ is close to zero ($\sim
0.0589$ at $a$) where the local slope should be close to the
maximum value. Near $\alpha \sim 0$, the local slope is the
product of the maximum amplitude of $\alpha$, say, $\alpha_0$ (in
the sinusoidal stretch $ab$) and $2\pi \nu$. (We have assumed
$\alpha = \alpha_0 \, sin 2\pi \nu t$ by dropping the phase
factor.) Thus, one should have $\dot \alpha \simeq 2 \pi \nu
\alpha_0$ when $\alpha \sim 0$. Using the value $\alpha_0 = 0.23$
from exact numerical solution and $\nu \sim 6.88$ at $F_{in} =
186.95$,  we find that $\dot \alpha \sim 10$ near $\alpha \sim 0$.
Indeed, this is satisfied only at $a$ where $\dot \alpha_{in} =
9.7 $. [Note that $\alpha$ is not symmetric around zero due to
the presence of $v$ in Eq. (\ref{FLOW1}) which has been ignored for
the purpose of present discussion.] However, $\dot \alpha_{in} =
18.3$ at $k$ is significantly higher than the slope permitted for
$\alpha$ to start a sinusoidal sojourn. This forces the orbit to
make one more small loop ($AB$ to $CD$ and back) so that the initial
value of $\dot \alpha_{in}$ is commensurate for $\alpha$ to start
a sinusoidal form. Indeed,  the initial values of $\dot \alpha$ at
all the earlier visits to $AB$ branch keep decreasing until it
reaches a value that is consistent to begin the sinusoidal growth.
Once this is satisfied, the monotonic increasing behavior  from
$a$ to $b$ is seen.  As we will show this is the mechanism
operating for high
$I$ and $V$ case.\\

\centerline {\bf Case III,  High $I$ and $V$}

For this case, even on the $AB$ branch, $\dot v/R$ cannot be ignored in
Eq. (\ref{genalpha}) and thus one needs to use coupled 
Eqs. (\ref{approxvdot}) and (\ref{genalpha}). Calculations follow much the
same lines and give correct values for $\Delta t $ and $\Delta F$
on both the branches during the rapid jumps.

Again, we need to answer when exactly does the system know to
switch from a rapid jumping mode to monotonically increasing on $AB$
or decreasing mode on $CD$?

Consider the last of the rapid jumps from $CD$ to $AB$ (just prior to
the point $a$) in Fig. \ref{fig5}(c).  The corresponding point in
the $\alpha$ plot [Fig. \ref{fig5}(b)] is shown on an expanded
scale in the inset. From this figure, it is clear that $\dot
\alpha$ has a positive slope at $k$, though of small magnitude
while at $a$, it has a value -9.7. The latter is close to the
natural (negative) slope of $\alpha$ when it begins the descending
branch of the  sinusoidal form. On the other hand, the slope of
$\alpha$ is positive at $k$ and hence will not allow the growth to
change over from a jumping mode to the sinusoidal growth form for
$\alpha$. One can note that the slopes at points of all the
earlier visits to AB [see Fig. \ref{fig5}(b) inset] keep
decreasing till the slope becomes negative required for the
monotonically decreasing trend of $\alpha$.  This is exactly the
same mechanism for $I = 10^{-2}$ and $V=1$ also, for the low $v$
branch, except that in this case, even the sign of the slope is
incompatible  for all the points prior to $a$ in Fig. \ref{fig5}
c. The mechanism operating on $CD$ (i.e., at the switching from
jumping mode to monotonically decreasing nature of $F$) is
essentially the same but arguments are a little more involved and
hence they are not presented.

Now, we consider the causes leading to the maximum and minimum
values taken by $F$ being much more than permitted by $f(v,V)$. As
this is dominant for $V=4$,  we illustrate this using Fig.
\ref{fig5}(a) and \ref{fig5}(c). We first note that Eq. (\ref{CONSTR}) constrains the
dynamically changing values of  $F(t)$ and $\alpha(t)$ to the
stationary  values of $f(v,V)$. Clearly, this implies that $F =
f(v,V)/( 1+ sin \alpha)$. A rough estimate of $F_{max}$ can be
obtained  by $F_{max} \sim f_{max}/(1 + \alpha_{min})$ with
$F_{min}$ determined by $\alpha_{max}$. This relation can be
easily verified by using the numerical values of $\alpha$. For
instance, for $V=4$ and $I = 10^{-2}$, $\alpha_{min} = -0.3$ and
$f_{max} = 307$. This gives $F_{max} = 438$ while the numerical
value from the phase plot for this case gives $433$ which is very
close. Similarly, using $\alpha_{max} = 0.62$ and $f_{min} = 293$,
we get $F_{min} = 181$ which compares well with the numerical
value of 180. We have verified this relation is respected for
various values of $V$ and $I$. For small $I$, $\alpha$ is  small,
we should not find much difference between $F_{max}$ ($F_{min}$)
and that of $f$.

\section{Summary and Conclusions}

We first summarize the results before making some relevant
remarks. We have carried out a study of the dynamics of an
adhesive roller tape using a differential-algebraic scheme used
for singular set of differential equations. The algorithm produces
stick-slip jumps across the two dissipative branches as a
consequence of the inherent dynamics. Our extensive simulations
show that the dynamics is much richer than anticipated earlier. In
particular the influence of inertia is shown to be dramatic. For
instance, even at low inertia, for small values of $V$, the
influence of inertia manifests with jumps of the orbit occurring even
before $F$ reaches $f_{max}$ (or $f_{min}$) which is quite
unexpected. More dominant is its influence for high
$I$ both for low $V$ and high $V$, though it is striking for the
latter case. Following the reasoning used in the PLC effect, we
introduce a dynamized curve $f(v,V)$ as resulting from competing
time scales of internal relaxation and imposed pull speed. The
modified peel force function leads to  the decreasing trend in the
 magnitude  of $\overline{\Delta F}$ with increasing pull
velocity, a feature observed in experiments. We have also
recaptured the essential features of the dynamics by a set of
approximations valid in different regimes of the parameter space.
These approximate solutions illustrate the influence of various
time scales such as that due to inertia, the elasticity of the
tape and that determined by the stationary peel force $f(v,V)$. We
also find the unusual canard type of solutions.

Here, it is worthwhile to comment  on the dynamical features of
the model. The numerical results themselves are too complex to
understand. A striking example of this is the series of force
drops seen on the descending branch of the pull force
[Fig. \ref{fig4}(c)]. This result is hard to understand as it would
amount to a partial relaxation of the pull force. However, a
partial relaxation is only possible in the presence of another
competing time scale (other than the imposed time scale). Another
example is the jumping of the orbit for low $I$ case,  from $AB$ to
$CB$ and vice versa even before the pull force reaches the extremum
values of $f(v,V)$.  For this reason, we have undertaken to make
this complex dynamics transparent using a set of approximations.
The basic idea here is to solve a single equation (or at most two
equations as in the high $I$ and $V$ case) which incorporates all
the relevant time scales. This method not only captures all the
results to within 10\% error but it also clearly brings out the
regimes of parameter space where these time scales become
important. This analysis also shows that the time scale due to
inertia of the roller tape shows up even for low $I$ which comes
as a surprise as one expects that for low inertia, the orbit
should stick to the stationary peel function. (Recall that for
low inertia, equations have been approximated by Lienard type of
equations by Maugis and Barquins \cite{MB}.) Our approximate
equations  demonstrate that a crucial role in inducing the jumps
{\it even at low inertia} is played by the high frequency
oscillations resulting from the inertia of the roller tape. As for
high inertia (both for low and high pull velocities), the time
scale due to inertia is responsible for the partial relaxation of
$F$ as shown.

A few comments may be in order on the bursting type of
oscillations in the peel velocity. Bursting type of oscillatory
behavior are commonly seen in neuro-biological systems \cite{CFL}.
Conventionally, bursting type oscillations arise in the presence
of homoclinic orbit~\cite{CFL}. Such bursting type of oscillations
have also been modeled using one dimensional map ~\cite{Rul}.
However, it is clear that the mechanism for bursting type of
oscillations in our case is different. In our case, this arises
due to the fact that the orbit is forced to jump between the
stable manifolds as a result of competing  time scale of inertia
and  the time scale for the evolution of $v$. (We note that the
latter itself  includes more than one time scale [see Eq.
(\ref{approxvdot})], namely  the contribution from the slopes of the
stable parts of the stationary curve $f(v,V)$ and that due to
elasticity of the tape. ) The bunching of the spikes in $v$ is the
result of $f(v,V)$ becoming flat for large $V$ and $I$. One other
comment relates to canard type solutions. Figure \ref{fig7} shows
one such solution. As mentioned, these type solutions arise from
sticking to the unstable manifold. In fact, a similar type of
solution is seen in Fig. \ref{fig5}(c). As noted earlier, all the
jumps from $CD$ to $AB$ or vice versa always occur when the peel
velocity reaches the limiting value where $f^{\prime}(v,V)=0$.
However, it can be seen from this figure, the orbit starting from
$c$ monotonically decreases well into the unstable part of $f$.
Thus, this solution also has the features of canards. It must be
stated that our approximate solutions cannot capture the behavior
of canards.

Finally, the results presented in this paper are on the nature of
dynamics  of  the model equations which so far had defied
solution. However,  comparison with experiments has been minimal
largely  due to the paucity of quantitative experimental findings
as stated earlier. Our analysis shows that the model
predicts periodic, saw-tooth \cite{MB}, as well as chaotic
solutions as reported in \cite{Gandur}. The high magnitude of the
Lyapunov exponents for the chaotic solutions in the low pull
velocities  is consistent with that reported earlier
\cite{Gandur}.  We note that the other quantitative experimental
feature reported by Refs.\cite{MB,Gandur} is the decreasing trend
of the average force drop magnitudes as a function of the pull
velocity  is also captured by our model (Fig. \ref{fig6}), a
result that  holds for both low and high inertia. This result is a
direct consequence of the dynamization of the peel force function,
i.e., dependence of the peel force on the pull velocity. We note
here that the complex dynamics at high velocities  (see Fig.
\ref{fig5}) is a direct result of the unstable part of dynamized
curve, $f(v,V)$, shrinking to zero. To the best of our knowledge,
this is first time  the result in Fig. \ref{fig6} has been
explained. As the hypothesis of dynamization captures the
decreasing trend of the force drops, it also suggests that the
underlying mechanism of competing time scales responsible for  the
peel force  depending on the pull velocity is likely to be correct
as in the PLC effect. Clearly, a rigorous derivation of the peel
force function from microscopic considerations that includes the
effect of the viscoelastic glue at the contact point is needed to
understand the dynamics appropriately.

\section{Acknowledgements}

The authors wish to thank A. S. Vasudeva Murthy of TIFR, Bangalore 
for useful discussions on DAE algorithm.
RD and AM wish to thank M. Bekele of Addis Ababa University, Ethiopia and M.
S. Bharathi of Brown Univ., USA for stimulating and friendly
discussions.  This work is financially supported by the 
Department of Science and Technology, New Delhi, India under the
grant SP/S2K-26/98.

\end{document}